\documentclass{article}
\usepackage{arxiv}
\usepackage[utf8]{inputenc} 
\usepackage[T1]{fontenc}    
\usepackage{hyperref}       
\usepackage{url}            
\usepackage{booktabs}       
\usepackage{cite}
\usepackage{amsmath,amssymb,amsfonts}
\usepackage{algorithmic}
\usepackage{graphicx}
\usepackage{textcomp}
\usepackage{xcolor}
\def\BibTeX{{\rm B\kern-.05em{\sc i\kern-.025em b}\kern-.08em
    T\kern-.1667em\lower.7ex\hbox{E}\kern-.125emX}}
\usepackage[misc,geometry]{ifsym} 
\usepackage[all]{xy}
\graphicspath{ {./images/} }

\begin{document}

\title{On The Fly Diffie Hellman for IoT}

\author{
  Jaime Díaz Arancibia \\
  Dept. Ciencias de la Computación e Informática\\
  Universidad de La Frontera \\
4811230 Temuco - Chile \\
  \texttt{jaimeignacio.diaz@ufrontera.cl} \\
   \And
 Vicente Ferrari
Smith\textsuperscript{ (\Letter)} \\
  Dept. Ciencias de la Computación e Informática\\
  Universidad de La Frontera \\
4811230 Temuco - Chile \\
  \texttt{v.ferrari01@ufromail.cl} \\
  \AND
  Julio López Fenner\textsuperscript{(\Letter)} \\
  Dept. Ciencias de la Computación e Informática\\
  Universidad de La Frontera \\
4811230 Temuco - Chile \\
  \texttt{jaimeignacio.diaz@ufrontera.cl} \\
}
\date{July 2019}
\maketitle

\begin{abstract}
The Internet of Things (IoT) is a fast growing field of devices being added to an interconnected environment in an abstract heterogeneous array of servers and other devices, called smart environments, ranging from private local (home) environments to nation-wide infrastructures, often accessible via unsecured wireless communications and information technologies, hence, massively open to attacks. In this paper we address some of issues that arise when connecting smart devices endowed with low computational capabilities to a home gateway via unsecured wireless communication channels, by using a One Time Pad (OTP) protocol based upon an On-the-fly Diffie-Hellman Key Exchange. Our assumptions are that only a user and the gateway have enough processing power to perform - say - secured RSA encrypted communication, hence relaxing the need for a trusted secure server outside the domain and that the protocol should at least be secure for a range of known attacks, as replay or DoS attacks.
\end{abstract}


\section{Introduction}
\label{sec:introduction}

Following \cite{kumar2015lightweight} and \cite{dey2019session}, the connection of an IoT device, endowed normally with limited processing power, to a central server using wireless networks requires following suitable authentication protocols and encryption, in order to avoid the well known Man in the Middle (MiM) attack, which we assume is a malicious third party having partial or complete access to the wireless communications between device and server. We  base our considerations mostly upon \cite{kumar2015lightweight} and the references cited therein.

For the system setup we propose establishing a secure connection between a smart device (SD) and the home gateway (HG) without referring to an external secure trusted server. Our protocol uses instead the End User as a One Time Trusted Server (OTTS) (wich is used only once at configuration time) for securely creating a Diffie-Hellman One Time Pad (OTP) for on-the-fly encryption between the device and the gateway. Once the setup protocol has been executed, all communications between the selected device and the server runs encrypted and can be hence considered secure. Communications between the server and the user are supposed to be secure enough as to be protected by RSA encryption and authentication. 

A potential drawback of this procedure could lie in an eventual loss, interruption or tampering of communications, which would require again user interaction for restarting the protocol, an issue which we address in section \ref{sec:reinit}.

This article proceeds as follows: Section \ref{sec:background} provides a brief description of the
basics of Diffie-Hellman key exchange and RSA authentication. Both RSA and DH require previous
agreement between parties (Alice and Bob) upon the modular arithmetic to be used, which normally
happens without concern about secrecy of those said parameters. If both parties have enough
computational power as to possess, each, valid RSA public and private keys, both algorithms can be
combined to securely agree upon key generation (KeyGen) for DH, which, nonetheless, still doesn't
avoid an active adversary to assume the role of one of the parties. Section \ref{sec:protocol}
presents our protocol which takes care of this possibility by adding a third party at setup time, the
user, with which the assumption of secure connection to a trusted external server can also be
relaxed. Section \ref{sec:reinit} deals with the issue of restarting the protocol. We close the
article with some comments upon the security analysis of the proposed protocol as well as a
discussion of its applicability to high throughput smart devices.
 
\section{Background}
\label{sec:background}

\subsection{Diffie-Hellmann Key Exchange}
The well known Diffie-Hellman Key Exchange \cite{diffie1976new} protocol proceeds as follows:
\begin{itemize}
    \item Alice and Bob share knowledge of two primes $g$ and $p$, in which $g$ is a primitive
    root\footnote{Recall that a
    primitive root $g$ of $p$ is a generator of the cyclic group $\mathcal{G}_p= \mathbb{Z}/ (p
    \mathbb{Z}) =
    \{ e, g^1, g^2, \dots , g^p\}$ and hence such that any invertible element has an inverse which is
    a power
    of $g$. } of $p$. 
    \item Alice generates or has a secret $a \in \{ 2,\dots p-2\}$ \cite{paar2009understanding}.
    \item Alice computes the quantity $x = g^a \mod p$ and sends $x$ to Bob (over the
    insecure channel).
    \item Bob knows $(g, p)$ and chooses his own secret $b$ for computing $y= g^b \mod p$, which is
    then sent over to Alice.
    \item Both Alice and Bob now compute $y^a \mod p$ and $x^b \mod p$ respectively.
    \item The computed result $K = y^a \mod p = x^b \mod p$ is identical for both, Alice and Bob,
    while all the transmitted data is indistinguishable from random, hence the insecure channel
    causes no trouble. $K$ can now be used for OTP Xor-encrypting exchanges between Alice and Bob.
\end{itemize}

\subsection{RSA Signature}

We consider now the standard RSA signature \cite{paar2009understanding} as follows:
Let Alice and Bob wish to share sensible information, which we denote by $Z$ (for example $Z$ could
be the pair ($g,p$) from the DH-Key Exchange described before). They proceed as follows:

\begin{itemize}
    \item Alice and Bob have each their private and public Key, which we denote by $K^A_{pub}$,
    $K^A_{priv}$,
    respectively $K^B_{pub}$, $K^B_{priv}$
    \item Alice sends Bob the triple ($Z, E_{K^A_{priv}} (Z)$)
    \item Bob decrypts Alice's message using her public Key and checks that $Z =
    D_{K^{A}_{pub}}(E_{K^A_{priv}}(Z))$. If not, the message has been intercepted and tampered with,
    else they
    are recognized as signed by Alice.
\end{itemize}

\subsection{Certificated agreement upon structural data }
Assume now that the structural data for establishing RSA authentication and DH should be agreed upon
by preserving privacy between parties Alice and Bob. A standard signature procedure allows for this
as follows: 

\begin{itemize}
    \item Assume Alice and Bob have each their own RSA public and private keys.
    \item Alice choose two primes $g$ and $p$ with $g$ a primitive root of $\mathcal{G}_p$, which
    will be her secret.
    \item She encrypts the pair ($g,p$) with her own private Key $K^A_{priv}$: $
    E_{K^A_{priv}}(g,p)$.
    \item She encrypts now the signed message $ \mathcal{M} = (g,p, E_{K^A_{priv}}(g,p))$ with Bob's
    public
    key: $E^B_{pub}(\mathcal{M})$ and sends it to Bob, which decodes it, check the signature using
    Alice's
    public key and - since the check will succeed - obtains as a result a valid shared ($g,p$) with
    which now a valid DH-Key Exchange can be safely performed between Alice and Bob.
\end{itemize}

Notice that the RSA authentication as described above is one-sided, in the following sense: The
structural space $\mathcal{G}_p$ is proposed by Alice and she keeps it secret. After completion
of the protocol, Bob thinks he knows for sure that Alice shared her secret $Z$ with him. But a
careful analysis shows that if Alice has been hacked, then a malicious Man-in-the-Middle (MiM), which
we call Daniel (Dan), may have intercepted Alice's message and replaced it with his own signed triple
($\tilde Z, E_{K^D_{priv}} (\tilde Z)$) after replacing Alice's public key with his own $K^D_{pub}$.
So now Bob will address Dan each time he thinks he is talking Alice to. This is a standard issue in
two-party authentication and is the reason why a handshaking protocol between Alice and Bob has to
happen as in the Station-to-Station (STS) protocol, see \cite{diffie1992authentication}, but also
explains variants of the kind
authenticated key agreement with key confirmation (AKC) \cite{blake1997key},  \cite{diffie1976new},
\cite{law2003efficient}.

\section{Protocol}
\label{sec:protocol}

By introducing the End User as a third party with enough processing power, we can relax the
assumption made upon Alice and establish a secure connection as follows:

Assume a given IoT device (which we denote by SD, or smart device) with low or limited computational
capability, which will take the role of Alice (like a temperature sensor, a door actuator or
similar), just powerful enough as to autonomously produce (pseudo) random numbers and perform
exponentiation in modular arithmetic. SD will be connected via an open unsecured channel (a wireless
connection, for example) to a home gateway (HG), which will be our Bob, which we assume is a server
(or computer) with higher computational power, at least enough as to be able to have its own public
and private key for RSA purposes. 

The scenario will be now the following: The End User, which we call U, proceeds to connect the SD
(Alice) to the HG (Bob) as in \cite{tschofenig2015architectural}, using for example a HAN protocol like ZigBee, known as SD-To-HG communication pattern. We assume that the user can write parameters onto the SD and aim to securely connect Alice to Bob by XOR-encrypting their communications via a One - Time - Pad established on-the-fly, as follows:

\subsection{System Setup}
In the following we use standard notation for encryption, decryption and concatenation as: 
\begin{table}[h]
    \centering
    \begin{tabular}{|c|l|} \hline
        Symbols & Description \\ \hline \hline
       $c=E_K(m)$  & Message $m$ is encrypted using secret key ($K$), \\ & producing the ciphertext $c$.\\ \hline 
       $m = D_K(c)$ & Message $m$ is decrypted using secret key ($K$) \\ & from the ciphertext $c$. \\ \hline
         $||$  &  String concatenation \\ \hline
    \end{tabular}
    \vspace{0.15cm}
    \caption{}
    \label{tab:my_label}
\end{table} 

Assuming RSA asymmetric key cryptography is available for the user U and the home gateway HG, as well as an offline interface for the User enabling him to write into the smart device SD parameters as needed, our system setup protocol proceeds as follows:

\begin{itemize}
    \item U generates a one-time (configuration session) RSA pair $K^{SD}_{pub}$, $K^{SD}_{priv}$ for Alice, as well as a pair $(g,p)$ with $p$ a prime number and $g$ a primitive root which defines the structural space $\mathcal{G}_p$.
    \item U writes into the SD the private key $K^{SD}_{priv}$.
    \item U sends to Bob (the gateway) the RSA encrypted message $E_{K^{HG}_{pub}}((g, p) || K^{SD}_{pub})$. \footnote{Strictly speaking, U should send the signed message encrypted with HG's publick key $E_{K^{HG}_{pub}}\left( ((g, p) \; || \; K^{SD}_{pub})\; || \; E_{K^U_{priv}}((g, p) \; || \; K^{SD}_{pub}) \right) $ allowing HG to verify U's identity, but this can be avoided if communication between HG and U is RSA protected or happens offline.}.
    \item Bob decrypts the message (if necessary authenticates Bob) and hence learns $(g,p)$ as well as $K^{SD}_{pub}$.
    \item Now Bob encrypts the pair $(g,p)$ together with his signature by using the RSA key he just learned:
    $E_{K^{SD}_{pub}}((g,p)|| E_{K^{HG}_{priv}}(g,p))$, and sends it over (the still insecure channel) to Alice.
    \item Alice requests Bob's public key and checks for message integrity (authentication) by verifying that  $D_{K^{HG}_{pub}}(E_{K^{HG}_{priv}}(g,p)) = (g,p)$.
\end{itemize}

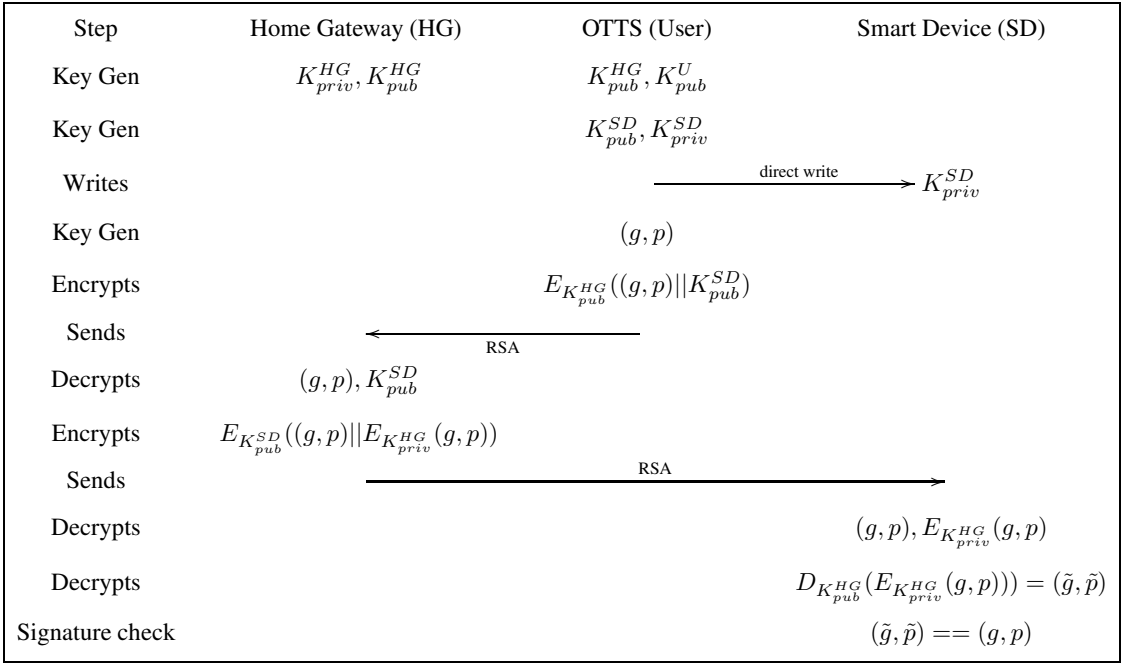
\begin{figure}[htbp]\center
    \resizebox{0.9\textwidth}{!}{\framebox[1\width]{
        $$\xymatrix@R=3pt@C=12pt{
        \text{Step} \hfill   & \text{Home Gateway (HG) }& \text{OTTS (User)} & \text{Smart Device (SD)}   \\         
           \text{Key Gen} & K^{HG}_{priv}, K^{HG}_{pub} & K^{HG}_{pub},K^{U}_{pub} &   \\
            \text{Key Gen} & & K^{SD}_{pub}, K^{SD}_{priv} &   \\
            \text{Writes} & & \ar[r]^{\text{direct write}} & K^{SD}_{priv}  \\
            \text{Key Gen} & & (g,p) & \\
            \text{Encrypts} & & E_{K^{HG}_{pub}}((g, p) || K^{SD}_{pub}) & \\
            \text{Sends} &  & \ar[l]^{\text{RSA}} \\
            \text{Decrypts} & (g, p), K^{SD}_{pub} & & \\
            \text{Encrypts} & E_{K^{SD}_{pub}}((g,p)|| E_{K^{HG}_{priv}}(g,p)) \\
            \text{Sends} & \ar[rr]^{\text{RSA}} & &   \\
            \text{Decrypts} & & & (g, p), E_{K^{HG}_{priv}}(g,p) \\
            \text{Decrypts} & & & D_{K^{HG}_{pub}}(E_{K^{HG}_{priv}}(g,p))) = (\Tilde{g}, \Tilde{p}) \\
            \text{Signature check} & & &  (\Tilde{g}, \Tilde{p}) == (g,p)  \\
        }$$
      }
    }
    \caption{System Setup for On-The-Fly DH}
    \label{fig:plusMP}
\end{figure}

As a result, Alice and Bob now share knowledge of $(g,p)$, provided securely by U and can hence Diffie-Hellman \textit{ad libitum} between them, which they will use for jointly creating the Key for the OTP. 

Observe that we deviate from the system design used in \cite{kumar2015lightweight} in that we do not assume the Service Provider (SP) as being secure (trusted) in the sense that, in our setting, generation and assignment of keys for securely handshaking between Alice and Bob is entrusted to the end user (U) by means of a one-time RSA authentication, using U and the SD on one side, and assuming secure communications between the HG and the User on the other. Also, requirements upon the SD are only in terms of performing basic modular arithmetic needed just for Diffie-Hellman.

\subsection{On the fly OTP}
Once Alice and Bob have securely identified themselves at system setup via the above procedure and shared the (now secret) values of $(g,p)$, they proceed to encrypt their communication via XOR-ing on the fly: Each time a packet of 
length $\sigma$ will be transmitted a DH procedure is performed in order to generate a new key with which the packet is XOR encoded. Hence the name 'on-the-fly'.  Via suitable padding and delay agreement, the protocol can be adapted to include the new (encoded) value of the key for the next data transfer, as well as the use of time stamps as Nonces in order to avoid replay attacks, see below.

Observe that unless Alice and/or Bob are compromised in the sense that, if after an attack either $K^{SD}_{priv}$ or $(g,p)$ have been changed or made public (see for example firmware attack \cite{liu2016smart}),  the OTP generated by sharing the key with DH is secure to replay attacks (by adequately using Nonces) and by itself, granting perfect security, since each time a packet is transmitted a new key is agreed upon between Alice and Bob.

\section{Re-initialization}
\label{sec:reinit}

Should communication between Alice and Bob be interrupted or fail the authentication, rebooting the system is easy if the private key of SD has not been compromised, it suffices to Bob to reestablish communication by creating a new pair $(g,p)$ and authenticate himself with Alice to share $(g,p)$ to her, provided - of course - that Bob has not been hacked into. If, on the contrary, $K^{SD}_{priv}$ has been exposed, user U has to manually re-initiate the system by furnishing a new system setup.

\section{Security analysis}
\label{sec:proof}

In this section we argue about the protocol's security properties (Authentication, message freshness, data integrity, etc) and its resistance to various attacks (Replay attack, Man in the Middle, etc).

\subsection{Authentication}

At the beginning of the protocol, as described by Section \ref{sec:protocol}-A, the User must have physical access to the SD device he is setting up, therefore, authentication between SD and U can be assumed as granted \textit{de facto}. On the other hand, authentication between U and HG proceeds through the RSA private and public key process, so that this part of the communication channel can be assumed secure: Any eavesdropper watching the traffic between U and HG will see, for example the sensible data $((g, p) || K^{SD}_{pub})$ encrypted with HG's public key $K^{HG}_{pub}$, so that only HG can decode it with his own private key $K^{HG}_{priv}$.

Similarly, once the User U has written the private key $K^{SD}_{priv}$ for SD onto the device, communication between SD and HG can be assumed authenticated via the same procedure as above.

\subsection{Key Establishment}

After authenticating each other, parties in the protocol will establish a DH-shared key each time a message has to be transmitted, which is then used to XOR-encrypt the said message, which is why we call it \textit{on-the-fly OTP} instead of \textit{session key}. In order for this to happen while preserving security, the size of the message should not exceed the size of the shared key, which is why, in principle, our approach would not be sufficient for, say, continuous data streaming as in video devices. Nonetheless, buffering via latency of the computations depending on technology could be brought in for inserting reestablishment of the DH-key session at fixed intervals, an issue which we do not pursue here and will be the subject of a forthcoming article.

Besides, as long as the structure data $(g, p)$  remains known only to the legitimate parties, communication between them via XOR-encrypted messages require only the generation of new private DH parameters for each message sent. This precludes a malicious party from gaining access to the next messages based upon knowledge of the previous ones, unless, of course, the SD device and its firmware can be completely simulated, together with the software being used to generate the keys. 

A procedure for dealing with this possibility can be easily derived along the lines of \cite{kumar2015lightweight}, section IV, authentication and key establishment. 

\subsection{Secrecy}

Before sending a message, the sending party (SD or HG) encrypts it using the shared key that must be at least of the same length as the message in order to use it as a OTP to generate the ciphertext.  By using the DH procedure, the parameters performing the DH-OTP will look random to any observer, which will be unable to determine the key, thus the transmitted XOR'd result will appear to be random to any eavesdropper or malicious party.

\subsection{Freshness}

This scheme generates a new (seemingly random) key every time a message should be sent. Even if the message is replayed, the ciphered text will not be recognized by the receiving party because of the different key they agreed upon before sending it. Hence replay attacks are naturally avoided with this procedure diminishing also the workload implicit in handshaking protocols including timestamps and the like.

\subsection{Masquerade or Forgery attack}
If a malicious attacker wanted to masquerade himself as the HG in order to read the sensitive data he would necessarily need to obtain first the private key $K^{HG}_{priv}$. Our protocol does certainly not exclude this from happening.

On the other hand, in order for an attacker to forge a valid message and send it to HG (in order to confuse or obfuscate the system by presenting intentionally wrong data, for example), the attacker should gain access to the structural data $(g, p)$, known in encrypted form by SD and HG only, in order to successfully simulate an authenticated SD message. For this, the intruder would need to physically gain access to the SD in the home network and thus extract this sensitive information, unless proper hardware security, (for example encrypting the device's flash memory) is applied, a resource that would have side effects in performance.

We do not address this issue here, but instead point out that it has been addressed elsewhere, see for example \cite{liu2016smart} and the references therein. 

\section{Conclusion}
\label{sec:conclusion}

Smart home environments are an essential part of the Internet of Things IoT, in which devices with low computational capabilities or power are connected to a communications net, accessible from 'the outside' through a Gateway HG, most of the cases in straight unsecured schemes or with unmodified or built-in or factory-default security parameters. 

Approaches for dealing with these kind of weakness involve often the assumption of a trusted external server. Yet more often than not, it is precisely inside the domains of the home environment that attackers can be successfully gain access to otherwise secured facilities. There are plenty of reports on security breaches that illustrates the importance of secure authentication and secure session keys even for the simplest smart devices and local home networks\footnote{See for example \url{https://www.iotforall.com/infamous-iot-hacks/} last retrieved july 2019. }

In this paper, we follow closely the protocol presented in \cite{kumar2015lightweight} and \cite{dey2019session} for Smart-Home-Environments but change some assumptions in both, the environment and the protocol itself, most notably in that the need for a trusted server outside the home network is no longer required. Instead, we replace the Trusted Server furnishing Keys and security by the (home) End User at system setup or configuration time, hence shifting the paradigm from a session-oriented protocol to an on-the-fly one.

Furthermore, by changing the assumptions upon the network architecture, we restrict ourselves strictly to the locality of the environment uncoupled or detached from any  external communication services. For this to succeed, we require a central device in the smart home environment with whom the user communicates, independently of the outside connections, that acts as Home Gateway. Equivalently, we require a HG with computational capabilities powerful enough as to provide full RSA encryption between the user and the HG.

For the time being, we do believe that our proposed on-the-fly protocol is a feasible option, at least for devices with very low bandwidth and low frequency of communications, granting privacy, authentication and security as described above, but also that via careful packet padding and/or buffering, the on-the-fly technique can also be applied for data streaming devices also. 

How this performs and is actually implemented is a subject whose discussion will be pursued in another work, with emphasis in performance.

\bibliographystyle{unsrt}
\bibliography{references.bib}  

\begin{thebibliography}{1}

\bibitem{kumar2015lightweight}
Pardeep Kumar, Andrei Gurtov, Jari Iinatti, Mika Ylianttila, and Mangal Sain.
\newblock Lightweight and secure session-key establishment scheme in smart home
  environments.
\newblock {\em IEEE Sensors Journal}, 16(1):254--264, 2015.

\bibitem{dey2019session}
Shreya Dey and Ashraf Hossain.
\newblock Session-key establishment and authentication in a smart home network
  using public key cryptography.
\newblock {\em IEEE Sensors Letters}, 3(4):1--4, 2019.

\bibitem{diffie1976new}
Whitfield Diffie and Martin Hellman.
\newblock New directions in cryptography.
\newblock {\em IEEE transactions on Information Theory}, 22(6):644--654, 1976.

\bibitem{paar2009understanding}
Christof Paar and Jan Pelzl.
\newblock {\em Understanding cryptography: a textbook for students and
  practitioners}.
\newblock Springer Science \& Business Media, 2009.

\bibitem{diffie1992authentication}
Whitfield Diffie, Paul~C Van~Oorschot, and Michael~J Wiener.
\newblock Authentication and authenticated key exchanges.
\newblock {\em Designs, Codes and cryptography}, 2(2):107--125, 1992.

\bibitem{blake1997key}
Simon Blake-Wilson, Don Johnson, and Alfred Menezes.
\newblock Key agreement protocols and their security analysis.
\newblock In {\em IMA international conference on cryptography and coding},
  pages 30--45. Springer, 1997.

\bibitem{law2003efficient}
Laurie Law, Alfred Menezes, Minghua Qu, Jerry Solinas, and Scott Vanstone.
\newblock An efficient protocol for authenticated key agreement.
\newblock {\em Designs, Codes and Cryptography}, 28(2):119--134, 2003.

\bibitem{tschofenig2015architectural}
Hannes Tschofenig, Jari Arkko, Dave Thaler, and D~McPherson.
\newblock Architectural considerations in smart object networking.
\newblock {\em RFC 7452}, 2015.

\bibitem{liu2016smart}
Jiajia Liu and Wen Sun.
\newblock Smart attacks against intelligent wearables in people-centric
  internet of things.
\newblock {\em IEEE Communications Magazine}, 54(12):44--49, 2016.

\end{thebibliography}

\end{document}